\documentclass[aps,pre,preprint]{revtex4}

\begin{document}

%\title{Response to Cohen and Mauzerall}
\title{Nonequilibrium work theorem for a system strongly coupled to a thermal environment}

\author{Chris~Jarzynski}
\affiliation{Theoretical Division, Los Alamos National Laboratory, NM 87545 USA}
\email{chrisj@lanl.gov}

\begin{abstract}
In a recent paper [{\it J.\ Stat.\ Mech.} P07006 (2004)],
E.G.D.\ Cohen and David Mauzerall (CM)
have argued that the derivation of the nonequilibrium work
relation given in [C.\ Jarzynski, {\it Phys.\ Rev.\ Lett.}\ {\bf 78}, 2690 (1997)]
is flawed.
Here I attempt to answer their criticisms,
both by presenting a detailed version of that derivation and by
addressing specific objections raised by CM.
The derivation presented here is in fact somewhat stronger than the one
I gave in 1997,
as it does not rely on the assumption of a weak coupling term between
the system of interest and its thermal environment.
\\ \\
LAUR-04-4712
\end{abstract}

\maketitle

In a recent paper\cite{cm},
Cohen and Mauzerall (CM) have raised questions about the validity of 
the nonequilibrium work relation, Eq.\ref{eq:nwt} below, which
relates the external work $W$ performed on a system during a nonequilibrium
process,
to the free energy difference $\Delta F$ between two equilibrium states of the system.
This prediction has been derived by various means in e.g.\ 
Refs.\cite{jar97a,jar97b,cro98,cro99,cro00,hum01,jar02,sun03,eva03,muk03},
and has been confirmed in an experiment
performed by Liphardt {\it et al}\cite{lip02},
along lines suggested by Hummer and Szabo\cite{hum01},
involving the forced unfolding and refolding of a single strand of RNA.\cite{sch03}
Two recent papers review theoretical, computational, and experimental aspects 
of the nonequilibrium work relation and related results.\cite{rit03,par04}
While most
of the criticisms of Ref.\cite{cm} are aimed at a derivation of Eq.\ref{eq:nwt} that I
published in 1997\cite{jar97a},
CM also suggest that other derivations 
(in particular that given by Crooks in Ref.\cite{cro98})
are relevant only for near-equilibrium processes, 
and that the experiment of Ref.\cite{lip02} cannot
be viewed as a confirmation of Eq.\ref{eq:nwt}.

Some of the issues raised by CM pertain to aspects of Eq.\ref{eq:nwt} that are
counter-intuitive, or at least sufficiently unusual
to arouse justifiable skepticism.
For instance, while the right side of Eq.\ref{eq:nwt} is familiar enough
-- it is a ratio of partition functions -- 
the left side is not.
The quantity inside the angular brackets is constructed by combining
two values -- the work $W$ performed on a system that is driven
out of equilibrium, and the temperature $T$ of the initial equilibrium
state of the system --
into something that looks like a Boltzmann factor,
namely $e^{-\beta W}$ (where $1/\beta = k_BT$).
To CM this is an {\it ad hoc}, unjustified construction, and 
I partially agree with them;
I can think of no good {\it a priori} reason to consider 
this particular quantity rather than some other.
However, as long as $\beta$ and $W$ are both well-defined,
then so is the value of $e^{-\beta W}$, and it is a perfectly
legitimate exercise to investigate its properties.
Any justification for embarking on such an investigation can only come
{\it a posteriori}:
if it leads us to an interesting and potentially useful result,
then that is sufficient reason for studying it in the first place!

My aim here is to address the arguments of CM,
in particular their assertion that the derivation
presented in Ref.\cite{jar97a} is flawed.
To do so, I will first present a detailed version of that derivation
(Section \ref{sec:derivation}),
with the goal of establishing the nonequilbrium work relation as a mathematical
identity, within the context of a Hamiltonian model for the evolution
of the system of interest and its thermal environment.
Following that, I will discuss several specific points raised by CM
(Section \ref{sec:points}).

\section{Derivation}
\label{sec:derivation}

The nonequilibrium work relation (or ``Jarzynski equality'', in CM) can be 
stated as follows.
Imagine a finite system that has been prepared in a state of equilibrium with 
a thermal environment at temperature $T$,
and suppose that we subject this system to a thermodynamic process,
by externally varying a work parameter of the system, $\lambda$,
from an initial value $A$ to a final value $B$.
In doing so we both drive the system out of equilibrium, and perform some amount
of work, $W$, on it.
The precise value of $W$ depends of course on the specific motions of the microscopic
degrees of freedom that constitute the system,
and these motions are in turn influenced by the degrees of freedom of the environment.
Therefore let us imagine that we carry out this process infinitely many times.
During each of these repetitions, or {\it realizations}, of the process,
we begin with the system and environment in a state of equilibrium,
and we always vary the work parameter $\lambda$
in precisely the same manner from $A$ to $B$.
After each realization we note down the amount of work $W$ performed on the
system during that realization, and in
the end we construct the distribution of work values, $\rho(W)$,
observed over this set of realizations of the process.
The nonequilibrium work relation states that this distribution satisfies a
strong constraint, namely:
\begin{equation}
\label{eq:nwt}
\Bigl\langle
e^{-\beta W}
\Bigr\rangle
\equiv
\int dW\,\rho(W)\,e^{\beta W}
=
e^{-\beta\Delta F},
\end{equation}
which remains valid regardless of how slowly or quickly we varied the work
parameter during the process.
Here, $1/\beta = k_BT$, and $\Delta F$ is the free energy difference between 
the equilibrium states associated with the initial and final values of the work parameter.
To be precise, let $Z_\lambda$ denote the partition function 
(defined by Eq.\ref{eq:Zlambda} below) corresponding to the
equilibrium state of the system of interest, when the work parameter is held
fixed at the value $\lambda$, and the system is in equilibrium with the environment.
Then the free energy of that equilibrium state is given by
the usual formula,
$F_\lambda = -\beta^{-1} \ln Z_\lambda$,
and the quantity $\Delta F$ appearing in Eq.\ref{eq:nwt} is {\it defined} to be
\begin{equation}
\label{eq:dfdef}
\Delta F \equiv F_B - F_A = -\beta^{-1}\ln\frac{Z_B}{Z_A}.
\end{equation}
A special case of Eq.\ref{eq:nwt}, pertaining to the situation in which
an external perturbation to the system is turned on and then off
(hence $\Delta F=0$) was derived earlier by Bochkov and Kuzovlev.\cite{bk}

I will now give a detailed version of the derivation of the nonequilibrium
work relation found in Ref.\cite{jar97a},
and will begin by spelling out the assumptions behind this derivation.

First, let us treat the system and its thermal environment
as a set of classical degrees of freedom that are well isolated from the rest of the
universe, and described by a Hamiltonian
\begin{equation}
\label{eq:hamiltonian}
{\cal H}(\Gamma;\lambda) =
H(x;\lambda) + H_E(y) + h_{\rm int}(x,y).
\end{equation}
Here, $x$ denotes a point in the phase space of the system of interest;
$y$ is a point in the (typically vastly larger)
phase space of the thermal environment;
$\Gamma=(x,y)$ is a point in the combined phase space of system and environment;
and $\lambda$ is an externally controlled work parameter.
The terms on the right side of Eq.\ref{eq:hamiltonian} correspond to
the bare Hamiltonian for the system of interest ($H$),
the bare Hamiltonian for the environment ($H_E$), and
the interaction energy between system and environment ($h_{\rm int}$).

Let us now imagine that we prepare the system and environment in an initial state of
thermal equilibrium at a temperature $T$, with the work parameter held fixed
at an initial value $\lambda=A$.
To be specific, suppose that this preparation is accomplished by placing
the combined system and environment in weak thermal contact with a much larger
``super-environment'' at temperature $T$, and then removing the super-environment
after an appropriate equilibration time.
As a result of this preparation, the system and environment find themselves in
a microstate $\Gamma_0=(x_0,y_0)$ that can effectively be viewed as being sampled
randomly from the canonical distribution in the full phase space:
\begin{equation}
\label{eq:fullcanon}
p(\Gamma_0) = \frac{1}{Y_A}\exp[-\beta {\cal H}(\Gamma_0;A)].
\end{equation}
The normalization factor $Y_A$ is a particular case ($\lambda=A$) of the quantity
\begin{equation}
\label{eq:calZ}
Y_\lambda \equiv
\int d\Gamma\,
\exp[-\beta{\cal H}(\Gamma;\lambda)].
\end{equation}
This is the classical partition function for the equilibrium state 
of the system and environment, when the work parameter is held
fixed at a value $\lambda$.

Having prepared the initial state of equilibrium and removed the
super-environment, we allow the system
and environment to evolve in time as we vary the
work parameter from $\lambda=A$ at $t=0$ to $\lambda=B$ at $t=\tau$,
according to some arbitrary but pre-determined schedule.
The microscopic history of system and environment during this process 
is described by a trajectory $\{\Gamma_t\}$
evolving under Hamilton's equations in the full phase space.
Here I use the notation $\{\Gamma_t\}$ to denote 
the {\it entire trajectory}, that is the microscopic history
of the system and environment from $t=0$ to $t=\tau$.
By contrast, the notation $\Gamma_t$ (without the braces) denotes simply
the microstate of the system and environment at a specific time $t$.
Similarly,
$\{\lambda_t\}$ will refer to the schedule for varying the work parameter
from $A$ to $B$, and $\lambda_t$ to the value of the work parameter
at a particular time $t$.
The schedule $\{\lambda_t\}$ specifies how we act on the system
during the process in question,
whereas $\{\Gamma_t\}$ specifies how the system and environment respond,
at the microscopic level, during a given realization of the process.

Let us interpret $H(x;\lambda)$ as the {\it internal energy} of the 
system of interest.\cite{qdefComment}
The net change in this quantity during a single realization
of the process is equal, identically, to
\begin{equation}
\label{eq:firstlaw}
H(x_\tau;B)-H(x_0;A) =
\int_0^\tau dt\, \dot\lambda \frac{\partial H}{\partial\lambda}(x_t,\lambda_t) +
\int_0^\tau dt\, \dot x \frac{\partial H}{\partial x}(x_t,\lambda_t).
\end{equation}
It is natural to interpret the first integral on the right side as the 
{\it external work} (or {\it mechanical work} in Ref.\cite{cm}) performed on the
system, and the second term as the {\it heat} absorbed by the system
(see for instance the discussions in Refs.\cite{cro98,sek98,jar98}).
Eq.\ref{eq:firstlaw} can then be viewed as a statement of the first law of
thermodynamics, which asserts
that the change in the internal energy of the system is due to
two contributions:
the work performed on the system,
\begin{equation}
\label{eq:wdef}
W \equiv \int_0^\tau dt\, \dot\lambda \frac{\partial H}{\partial\lambda}(x_t,\lambda_t),
\end{equation}
and the heat absorbed by the system,
\begin{equation}
\label{eq:qdef}
Q \equiv \int_0^\tau dt\, \dot x \frac{\partial H}{\partial x}(x_t,\lambda_t).
\end{equation}

Eqs.\ref{eq:wdef} and \ref{eq:qdef} define
$W$ and $Q$ in terms of the microscopic history of the system alone, $\{x_t\}$.
In effect, while Nature integrates Hamilton's equations in the full phase
space, we observers need to monitor only the degrees of freedom of the system
of interest in order to deduce how
much work was performed on the system, and how much heat was absorbed by it,
during a given realization of the process.
Note, however, that for a realization described by a trajectory $\{\Gamma_t\}$
in the full phase space, we have:\cite{explain1}
\begin{eqnarray}
{\cal H}(\Gamma_\tau;B) - {\cal H}(\Gamma_0;A) &=&
\int_0^\tau dt\,\frac{d}{dt}{\cal H}(\Gamma_t;\lambda_t) \\
&=&
\int_0^\tau dt\,\dot\lambda\frac{\partial{\cal H}}{\partial \lambda}(\Gamma_t;\lambda_t) \\
&=&
\int_0^\tau dt\,\dot\lambda\frac{\partial H}{\partial \lambda}(x_t;\lambda_t),
\end{eqnarray}
i.e.
\begin{equation}
\label{eq:wdH}
W = {\cal H}(\Gamma_\tau;B) - {\cal H}(\Gamma_0;A).
\end{equation}
This tells us that the work performed on the system of interest is
equal to the net change in the Hamiltonian of the combined system and environment.

The distinction between Eqs.\ref{eq:wdef} and \ref{eq:wdH} is an important one:
the former {\it defines} the external work performed on the system,
in terms of its microscopic evolution;
the latter states that 
the quantity $W$ thus defined
is equal to the net change in the combined energy of the system and environment
(under the assumption of Hamiltonian evolution for the combined system and environment).

The preceding paragraphs have specified a {\it model} used
to represent a system that is both in contact with a thermal environment,
and subject to an externally controlled work parameter.
Two strong but commonly made assumptions underlie this model:
first, that quantum effects can be ignored;
second, that the system and environment can be treated as being isolated from
all other degrees of freedom.
I will now show that Eq.\ref{eq:nwt} follows as a direct consequence
of this model.

So far the discussion has focused on
a single realization of the thermodynamic process in question.
Now imagine that we carry out the process very many 
-- in principle, infinitely many -- times.
We always prepare the system and environment in equilibrium as described above, 
and we always use the same schedule $\{\lambda_t\}$ to vary
the work parameter from $A$ to $B$.
In other words, we act on the system in precisely the same way, over and over again.
Nevertheless, the microscopic history of the system and environment, $\{\Gamma_t\}$,
will differ from one realization to the next, simply because the initial microstate
$\Gamma_0$ differs from one realization to the next (see Eq.\ref{eq:fullcanon}).
Over the course of each realization we observe the evolution of the system's 
degrees of freedom, $\{x_t\}$,
and from that empirical data we compute the value of $W$ using Eq.\ref{eq:wdef}.
Finally, from the set of work values collected over these realizations,
we construct the average of $e^{-\beta W}$.
Formally,
\begin{equation}
\label{eq:expav_emp}
\Bigl\langle
e^{-\beta W}
\Bigr\rangle =
\lim_{N\rightarrow\infty}
\frac{1}{N}
\sum_{n=1}^N
e^{-\beta W_n},
\end{equation}
where $N$ denotes the number of realizations, and $W_n$ is the work performed on
the system during the $n$th realization.
As above, $1/\beta=k_BT$, where $T$ is the initial temperature at which 
the system and environment are prepared, which is ``the only known temperature available'', 
as CM correctly point at the bottom of page 4 of Ref.\cite{cm}.

The preceding paragraph describes how to construct the desired average
from a series of measurements.
Let us now analyze this quantity theoretically:
for a given 
Hamiltonian ${\cal H}$, initial temperature $T$, and schedule $\{\lambda_t\}$,
what value will we obtain for the average defined by Eq.\ref{eq:expav_emp}?
Note that this problem is fully specified:
the probability distribution for the initial conditions $\Gamma_0$
is given by Eq.\ref{eq:fullcanon};
the subsequent evolution in the full phase space, $\{\Gamma_t\}$, 
is determined by Hamilton's equations;
and the quantities $\beta $ and $W$ are precisely defined.
Thus the question we have just posed has a unique answer within the context
of our model, and it remains only to do the math.

To carry out the analysis, let us introduce a function
$\tilde W(\Gamma_0)$, which is the work performed on the system
for a realization launched from initial conditions $\Gamma_0$ in the full
phase space.
(The initial conditions $\Gamma_0$ uniquely determine a trajectory $\{\Gamma_t\}$
in the full phase space,
and from such a trajectory the value of $W$ can be obtained through either
Eq.\ref{eq:wdef} or Eq.\ref{eq:wdH}.)
Then the average we wish to evaluate can be written as
\begin{equation}
\Bigl\langle
e^{-\beta W}
\Bigr\rangle =
\int d\Gamma_0\,
p(\Gamma_0)\,
\exp[-\beta\tilde W(\Gamma_0)],
\end{equation}
with $p(\Gamma_0)$ as given by Eq.\ref{eq:fullcanon}.
If we now combine Eq.\ref{eq:fullcanon} with Eq.\ref{eq:wdH}, then after
a single cancellation we get
\begin{equation}
\label{eq:almost_almost_there}
\Bigl\langle
e^{-\beta W}
\Bigr\rangle =
\int d\Gamma_0\,
\frac{1}{Y_A}\,
\exp[-\beta{\cal H}(\Gamma_\tau(\Gamma_0);B)],
\end{equation}
where the notation stresses the fact that Hamiltonian evolution is
deterministic:
$\Gamma_\tau(\Gamma_0)$ represents the final microstate in the full phase space,
for the realization launched from the initial microstate $\Gamma_0$.
Since there is a one-to-one correspondence between
the initial conditions $\Gamma_0$ and the final conditions $\Gamma_\tau$,
we can perform a change of variables in the above integral:
\begin{equation}
\int d\Gamma_0 \cdots \quad =\quad
\int d\Gamma_\tau \,
\Biggl\vert \frac{\partial\Gamma_\tau}{\partial\Gamma_0} \Biggr\vert^{-1}
\cdots,
\end{equation}
where $\vert\partial\Gamma_\tau/\partial\Gamma_0\vert$ is the Jacobian associated
with this change of variables.
By Liouville's theorem, this Jacobian is equal to unity.\cite{liouville}
We therefore have
\begin{equation}
\label{eq:almost_there}
\Bigl\langle
e^{-\beta W}
\Bigr\rangle =
\int d\Gamma_\tau\,
\frac{1}{Y_A}\,
\exp[-\beta{\cal H}(\Gamma_\tau;B)]
=\frac{Y_B}{Y_A},
\end{equation}
by Eq.\ref{eq:calZ}.
Since the ratio $Y_B/Y_A$ depends only on the parameter values $\lambda=A$ and $B$,
and on the temperature $T$,
Eq.\ref{eq:almost_there} already establishes a strong result.
Namely, even though the distribution of work values $\rho(W)$ generally depends
on the specific protocol for varying $\lambda$ from $A$ to $B$,
the quantity $\int dW\,\rho(W)\,e^{-\beta W}$ does not!

To this point, no approximations have been made in the analysis.
Now, the right side of Eq.\ref{eq:almost_there} is a ratio of partition
functions of the combined system {\it and environment} (see Eq.\ref{eq:calZ}).
We want to replace this by some expression pertaining to the system itself.
In Ref.\cite{jar97a} this was accomplished by explicitly assuming the
interaction energy $h_{\rm int}$ to be negligible in comparison with
the other two terms in the Hamiltonian ${\cal H}$.\cite{weakCoupling}
In many situations of physical interest, however, $h_{\rm int}$ is {\it not}
negligibly small.
In this case the evaluation of the ratio $Y_B/Y_A$ requires
a bit more effort.
As a starting point, let us recall that when $h_{\rm int}$ is finite,
the equilibrium distribution of the system of interest is given by the following
modification of the familiar Boltzmann-Gibbs formula:
\begin{equation}
\label{eq:modBoltz}
p_S(x;\lambda) \propto \exp[-\beta H^*(x;\lambda)],
\end{equation}
where 
\begin{equation}
\label{eq:Hstar}
H^*(x;\lambda) \equiv
H(x;\lambda) -\beta^{-1}\ln
\frac{\int dy \exp\Bigl[-\beta[H_E(y)+h_{\rm int}(x,y)]\Bigr]}
     {\int dy \exp [-\beta H_E(y)]}
\end{equation}
is a {\it potential of mean force} (PMF) associated with the
phase space variables of the system of interest.\cite{kir35}
The notation $p_S$ indicates the probability
distribution of the system of interest;
this is obtained from the probability distribution in the full phase space,
Eq.\ref{eq:fullcanon}, by integrating over the environmental degrees of freedom.\cite{rou01}
For the equilibrium distribution given by Eq.\ref{eq:modBoltz}, the partition function 
(normalization factor) is 
\begin{equation}
\label{eq:Zlambda}
Z_\lambda = \int dx\,\exp[-\beta H^*(x;\lambda)].
\end{equation}
With these definitions, we have
\begin{equation}
\label{eq:Yfactorization}
Y_\lambda = Z_\lambda\cdot
\int dy\,\exp[-\beta H_E(y)],
\end{equation}
which immediately implies
\begin{equation}
\frac{Y_B}{Y_A} = \frac{Z_B}{Z_A}. 
\end{equation}
Eq.\ref{eq:Yfactorization} is not an approximation, but follows identically from 
Eqs.\ref{eq:calZ}, \ref{eq:Hstar}, and \ref{eq:Zlambda}
along with the definition of ${\cal H}$, Eq.\ref{eq:hamiltonian}.
Combining these results with Eqs.\ref{eq:dfdef} and \ref{eq:almost_there}, 
we finally arrive at the nonequilibrium work relation:
\begin{equation}
\label{eq:nwt_finally}
\Bigl\langle
e^{-\beta W}
\Bigr\rangle
= \frac{Y_B}{Y_A}
= \frac{Z_B}{Z_A}
= e^{-\beta\Delta F},
\end{equation}
{\it without resorting to a weak-coupling assumption}.
For physical situations in which $h_{\rm int}$ happens to be negligible, 
we recover the situation discussed in Ref.\cite{jar97a}.

Note that the quantity $\Delta F$ has been defined {\it mathematically},
in terms of a ratio of partition functions (Eqs.\ref{eq:dfdef},\ref{eq:Zlambda}).
In the Appendix, I briefly argue that it is reasonable to view
this quantity as a {\it physical} free energy difference.

It is important to stress that the derivation which has just been presented
is {\it exact}: Eq.\ref{eq:nwt_finally} is a mathematical equality,
given the model specified above.
The key feature of this model is that the system and environment are treated
as an isolated collection of classical degrees of freedom evolving under 
Hamilton's equations.
While this model represents the traditional approach of classical
statistical mechanics, there is a subtlety associated with it, even if we agree to
ignore quantum effects.
This subtlety arises from the fact that it is in practice impossible to completely
isolate a system and its immediate thermal environment;
unavoidable interactions with the rest of the universe
introduce effectively random perturbations to the
evolution of the trajectory $\{\Gamma_t\}$.
(Moreover these perturbations are correlated with $\{\Gamma_t\}$.)
No matter how weak these perturbations may be, their effect
on a given trajectory becomes magnified exponentially with time
if the evolution of $\{\Gamma_t\}$ is chaotic, as is typically the
case for a realistic thermal environment.
With this in mind, are we really justified in invoking Liouville's theorem
in going from Eq.\ref{eq:almost_almost_there} to Eq.\ref{eq:almost_there}?
This theorem reflects a delicate balance between sets of initial and final conditions
of trajectories evolving under Hamilton's equations,
a balance that might well be upset by the addition of even the tiniest
amount of randomness into the equations of motion.
Questions such as this make it all the more important that Eq.\ref{eq:nwt} be
tested in actual laboratory experiments.

The above derivation relies explicitly on the assumption that the initial
equilibrium state 
is represented by a canonical distribution in the full phase space
(Eq.\ref{eq:fullcanon}).
This assumption was justified by the somewhat artificial construct of a
super-environment.
However, the validity of Eq.\ref{eq:nwt} might not depend as strongly on
a literal interpretation of Eq.\ref{eq:fullcanon} as the derivation suggests.
For instance, in the pulling experiment of Ref.\cite{lip02},
the microscopic system of interest -- a single strand of RNA, two
micron-size beads, and the DNA handles used to attached the RNA to the beads --
is immersed in a macroscopic bath of water molecules.
As long as that bath is prepared at a well defined temperature,
one intuitively expects that the behavior of the biomolecule,
immersed deep within the aqueous solution,
will not depend in any significant way on whether the combined system and
environment are prepared {\it exactly} in the canonical distribution
given by Eq.\ref{eq:fullcanon}.
Thus we would expect the same outcome (apart from extremely small corrections)
if the initial conditions $\Gamma_0$ were instead sampled from a
microcanonical distribution, $p\propto\delta(E-{\cal H})$;
this is a variant of the usual expectation of {\it equivalence of ensembles}.\cite{rue99}
For a quantitative discussion of this issue, see Section II.B of
Park and Schulten.\cite{par04}

\section{Response to specific points}
\label{sec:points}

A central claim of Ref.\cite{cm} is that the heat exchange between 
the system and environment has ``not been properly taken into account'' (page 5 of Ref.\cite{cm}.)
While it is true that the derivation given above never makes explicit use of 
the quantity $Q$,
this does not imply that $Q$ is assumed to be zero,
or that in some other way the heat has been mishandled in the analysis.
It simply means that -- within the context of the model --
one can evaluate the average of $\exp(-\beta W)$ without mentioning $Q$ in the calculation.

As an example of their claim that heat has not been treated properly,
CM consider the situation in which 
the system of interest is out of equilibrium at the moment when 
the work parameter reaches its final value (page 5 of Ref.\cite{cm}).
It is worthwhile to discuss this example in some detail.
To begin, recall that 
$\Delta F$ should always be understood as
the free energy difference between {\it the two equilibrium states associated with the
initial and final values of the work parameter},
rather than as the free energy difference between the initial and final states of 
the system of interest.
(Indeed, if the system is out of equilibrium at the end of the process, then its final free
energy might not be well-defined.)
To be absolutely precise,
for any process during which the work parameter is varied from $A$ to $B$,
the quantity $\Delta F$ appearing in Eq.\ref{eq:nwt} is defined by
(see Eqs.\ref{eq:dfdef},\ref{eq:Zlambda})
\begin{equation}
\label{eq:dF_explicit}
\Delta F = 
\frac{\int dx\,\exp[-\beta H^*(x;B)]}
     {\int dx\,\exp[-\beta H^*(x;A)]},
\end{equation}
regardless of whether or not the system is in equilibrium at the end of the process.

Now let us consider a two-stage schedule for varying the work parameter.
During the first stage ($0\le t\le\tau_1$),
$\lambda$ is changed in some arbitrary way from $A$ to $B$;
during the second stage ($\tau_1\le t\le\tau_2$),
$\lambda$ is held fixed at the value $B$.
Let us assume that the system is out of equilibrium at the time $\tau_1$,
but that the second (``relaxation'') stage is sufficiently long for the system to
relax to equilibrium.
Now imagine two observers, one of whom monitors the behavior of the system only during
the first stage,
while the other monitors the behavior during both stages.
Thus according to the first observer the process ends at time $\tau_1$
(with the system out of equilibrium),
whereas the second observer contends that the process ends at a later time $\tau_2$
(with the system in equilibrium).
As before, we imagine infinitely many repetitions of the process.

For every realization, the two observers agree on the precise amount of work performed
on the system during the process, even though they disagree as to when the process ends.
This follows from Eq.\ref{eq:wdef}: since $\lambda$ is fixed for $t>\tau_1$,
there is no contribution to $W$ from the relaxation stage.
Therefore, when the two observers independently construct the average 
$\langle\exp(-\beta W)\rangle$
after many realizations of the process, they both arrive at the same number
for the left side of Eq.\ref{eq:nwt}.
Moreover, since both observers agree that the work parameter begins at $A$ (at $t=0$)
and ends at $B$ (at $t=\tau_1$ or $t=\tau_2$),
they also agree on the value of $\Delta F$, as defined by Eq.\ref{eq:dF_explicit}.
Hence when using their data to assess the validity of Eq.\ref{eq:nwt} the two
observers will be comparing exactly the same numbers.
In other words,
whether or not we choose to include a relaxation stage -- during
which we hold $\lambda$ fixed at its final value, so as to let the system come
to equilibrium --
has no bearing whatsoever on the validity of Eq.\ref{eq:nwt}.
This is a simple consequence of the definitions
of the quantities $W$ and $\Delta F$.

Although no work is done on the system during the relaxation stage,
there is typically a certain amount of heat exchanged between 
system and environment during this stage.
Thus the observed values of $Q$ generally do depend on whether the 
process is defined to end at time $\tau_1$ or time $\tau_2$.
But this in no way affects the validity of Eq.\ref{eq:nwt}, since that prediction
concerns work, as defined by Eq.\ref{eq:wdef}, and not heat.

Another point raised by CM concerns the factor $\beta=1/k_BT$,
where $T$ denotes the initial temperature of the system and environment.
Once the system is driven away from equilibrium, it might not
have a well-defined temperature, and even if it did there is no guarantee that it
would be equal to the initial temperature $T$.
Therefore, in the expression $e^{-\beta W}$, a value that pertains to a system
out of equilibrium ($W$) is divided by a temperature ($T$) that does not meaningfully
represent the state of that system, except at $t=0$.
CM assert, with some justification, that such a pairing appears arbitrary and 
without foundation (see e.g.\ page 4 of Ref.\cite{cm}).
Note the nature of this criticism:
CM do not claim that the quantity $e^{-\beta W}$ is somehow
inherently ``illegal'' or ill-defined, but rather that it is {\it ad hoc}.
However, as discussed briefly in the introduction above, 
as long as $\beta$ and $W$ are well-defined,
it is perfectly acceptable to inquire about the average of $e^{-\beta W}$
over different realizations of the process.
As the detailed calculation of Section \ref{sec:derivation} reveals in the context
of a particular model --
and as has been shown by a number of other derivations using significantly different
models\cite{jar97b,cro98,cro99,cro00,hum01,jar02,sun03,eva03,muk03} --
this average works out to be equal to $e^{-\beta\Delta F}$.
It might be surprising that a construction as intuitively unnatural as
$e^{-\beta W}$ should lead to such a simple result,
but this does not automatically invalidate the result.
Indeed, the fact that this quantity seems unnatural might simply reflect 
our limited intuition regarding nonequilibrium processes.\cite{artur}

In Section 2 of their paper, CM consider two factors that play an important role 
in determining whether 
a process is reversible or irreversible.
The first is the rate of heat transfer between the system and environment, $\dot c$,
which is related to the strength of the coupling between them;
the second is the rate at which work is performed, 
$\dot w = \dot\lambda\,\partial H/\partial\lambda$.
CM discuss several cases illustrating how the balance between
$\dot c$ and $\dot w$ affects the reversibility or irreversibility of the process.
Their discussion is physically motivated and certainly seems correct,
but it does not bear on the validity
of the derivation of the nonequilibrium work relation given in Ref.\cite{jar97a}.
As shown in detail in Section \ref{sec:derivation} above,
that derivation is based on very general considerations involving Hamilton's equations, 
Liuoville's theorem, and the use of the canonical ensemble to represent the initial 
equilibrium state of the system and environment.
These considerations remain valid independently of the values of $\dot c$ and $\dot w$.
To put it another way:
the analysis presented in Section \ref{sec:derivation} above
depends neither on the rate at which the work parameter is varied,
nor on the strength of the coupling between the system and its
environment, hence it
is as valid for irreversible processes as it is for reversible ones.

In Section 5 of Ref.\cite{cm} CM discuss the example of an
isolated harmonic oscillator whose frequency is externally switched 
from an initial value $\omega_0$ to a final value $\omega_1$
(where $\omega_1>\omega_0$),
over a switching time $t_s$.
In Ref.\cite{jar97b} I had shown that
in the limit of infinitely slow switching,
and assuming a canonical distribution of initial conditions for the oscillator,
one can solve exactly for the distribution of work values:
\begin{equation}
\label{eq:ho}
\lim_{t_s\rightarrow\infty} \rho(W) =
\frac{\omega_0\beta}{\omega_1-\omega_0}
\exp\Biggl(
\frac{-\omega_0\beta W}{\omega_1-\omega_0}
\Biggr)
\theta(W),
\end{equation}
where $\theta(\cdot)$ is the unit step function.
It is easy to verify that this distribution satisfies the nonequilibrium
work relation, Eq.\ref{eq:nwt} above.
The derivation of Eq.\ref{eq:ho} makes use
of an adiabatic invariant and therefore is
valid only in the limit $t_s\rightarrow\infty$.
This does not, however, imply that the nonequilibrium work relation
fails for finite values of $t_s$,
only that for finite switching times it is not easy to obtain an
exact expression for $\rho(W)$.
(Exactly solvable models are hard to come by!
See, however, Ref.\cite{mazonka}.)
Therefore in Ref.\cite{jar97b} the analytical calculation of $\rho(W)$
for infinite switching times (Eq.\ref{eq:ho} above) was supplemented
by numerical experiments carried out for five different finite values
of $t_s$.
The results of these experiments were in excellent agreement with
Eq.\ref{eq:nwt}, as shown by the diamonds in Fig.2 of Ref.\cite{jar97b}.
It is difficult to reconcile these results with CM's statement that
the nonequilibrium work relation ``is critically dependent on adiabatic
invariance for finite $t_s$, even for the harmonic oscillator'' (page 14),
particularly since in Ref.\cite{jar97b} adiabatic invariance was only 
invoked in the limit $t_s\rightarrow\infty$.

Of course a single example does not establish universal validity.
General proofs of Eq.\ref{eq:nwt} for Hamiltonian systems were given
in Refs.\cite{jar97a} and \cite{jar97b}, and the harmonic oscillator was meant to
serve only as a simple illustration.

It is a pleasure to acknowledge useful discussions and correspondence with
A.\ Adib, C.\ Bustamante, E.G.D.\ Cohen, G.E.\ Crooks, J.\ Jarzynski,
J.\ Liphardt, and F.\ Ritort.
This research was supported by  the Department of Energy,  under contract W-7405-ENG-36.

\section*{Appendix}

Eq.\ref{eq:Yfactorization} can be written explicitly as
\begin{equation}
\int d\Gamma\,e^{-\beta{\cal H}(\Gamma;\lambda)} =
\int dx\,e^{-\beta H^*(x;\lambda)}
\,\cdot\,
\int dy\,e^{-\beta H_E(y)}.
\end{equation}
Thus the partition function for the combined system and environment
factorizes nicely as the product of two partition functions, one for the
system of interest (which includes all the effects of the interaction
energy) and the other for the environment.
If we take the natural logarithm of both sides and multiply by $-\beta^{-1}$,
we can rewrite the above result as
\begin{equation}
{\cal F}_\lambda = F_\lambda + F_E^0,
\end{equation}
where ${\cal F}_\lambda = -\beta^{-1}\ln Y_\lambda$ can be viewed as 
the equilibrium free energy of the combined system and environment,
and $F_E^0 = -\beta^{-1} \ln \int dy\,e^{-\beta H_E(y)}$
as that of the bare environment.
Note that $F_E^0$ is a macroscopic quantity, describing a macroscopic
thermal environment,
whereas the characteristic magnitude of $F_\lambda$ is determined by
the size of the system of interest;
e.g.\ $F_\lambda$ (and therefore $\Delta F$)
is microscopic for a single-molecule pulling experiment.

Given the definitions in Section \ref{sec:derivation},
it is easy to show that the quantity $F_\lambda = -\beta^{-1}\ln Z_\lambda$
satisfies
$\partial F_\lambda/\partial\lambda = \langle \partial H^*/\partial\lambda\rangle_\lambda^{\rm eq}
= \langle \partial H/\partial\lambda\rangle_\lambda^{\rm eq}$,
or equivalently
\begin{equation}
\label{eq:ti}
\Delta F =
\int_A^B d\lambda\,\Biggl\langle
\frac{\partial H}{\partial\lambda}
\Biggr\rangle_\lambda^{\rm eq} \quad .
\end{equation}
Here $\langle\cdots\rangle_\lambda^{\rm eq} = \int dx\,p_S(x;\lambda)\cdots$ 
denotes an equilibrium average at a fixed value of the work parameter.
(The derivation of Eq.\ref{eq:ti}, not reproduced here, is just a few lines long,
and essentially identical to the derivation of the well-known {\it thermodynamic
integration} identity; see e.g.\ Ref.\cite{frenkel-smit}.)

Now suppose we carry out a {\it reversible} process:
the system passes through a continuous sequence of equilibrium states
as we slowly vary $\lambda$ from $A$ to $B$.
Thus, at any time $t$ during this process, the system of interest
is sampling its phase space according to the equilibrium distribution
($p_S$) corresponding to the current value of work parameter, $\lambda_t$.
In this situation we can replace the value of $\partial H/\partial\lambda$
appearing in the definition of work (Eq.\ref{eq:wdef}), by its equilibrium
average:
\begin{equation}
W 
\rightarrow
\int_0^\tau dt\, \dot\lambda
\Biggl\langle
\frac{\partial H}{\partial\lambda}
\Biggr\rangle_{\lambda_t}^{\rm eq} 
=
\int_A^B d\lambda\,
\Biggl\langle
\frac{\partial H}{\partial\lambda}
\Biggr\rangle_\lambda^{\rm eq} = \Delta F,
\end{equation}
invoking Eq.\ref{eq:ti}.
The arrow denotes that we are considering the special case of a reversible,
quasi-static process.
By these arguments, then, $W=\Delta F$
for any reversible process during which $\lambda$ is changed, quasi-statically,
from $A$ to $B$.
This suggests that we are justified in interpreting $\Delta F$,
defined mathematically in terms of the
modified Boltzmann distribution, Eq.\ref{eq:modBoltz},
as a physical equilibrium free energy difference.


\begin{thebibliography}{99}
\bibitem{qdefComment}
This is a reasonable interpretation when the
coupling between system and environment is negligibly weak.
However, when this is not the case, then the system and the environment
effectively share the energy in the term $h_{\rm int}$, and it is
not completely obvious whether this energy should be viewed as belonging
to the system, or to the environment.
In that situation, other considerations (not discussed here) suggest that
the quantity $H^*(x;\lambda)$ (Eq.\ref{eq:Hstar}) should replace
$H(x;\lambda)$ in Eqs.\ref{eq:wdef} and \ref{eq:qdef}.
This will change the definition of heat, Eq.\ref{eq:qdef},
but will have no effect on the definition of work, Eq.\ref{eq:wdef},
since $\partial H^*/\partial\lambda = \partial H/\partial\lambda$.
Hence whether we use $H$ or $H^*$ in Eqs.\ref{eq:wdef} and \ref{eq:qdef},
has no bearing on the derivation of Eq.\ref{eq:nwt} presented in the present
section.
\bibitem{liouville}
Imagine an infinitesimal cell of volume $\epsilon$ centered around the
point $\Gamma_0$ in the full phase space.
If we propagate a trajectory for a time $\tau$ from each set of initial
conditions found within this cell, then
we will end up with an infinitesimal cell of final conditions,
of volume $\epsilon^\prime$,
containing the point $\Gamma_\tau$.
The Jacobian, by definition, is the ratio of the volumes of these two cells,
$\epsilon^\prime/\epsilon$.
But Liouville's theorem tells us that Hamiltonian dynamics gives rise to no
phase space expansion or contraction, hence $\epsilon^\prime=\epsilon$.
\bibitem{weakCoupling}
In this situation the contributions to $Y_A$ and $Y_B$ from 
the environmental degrees of freedom cancel one another:
\begin{equation}
\frac{Y_B}{Y_A} =
\frac{\int d\Gamma\, \exp[-\beta{\cal H}(\Gamma;B)]}
     {\int d\Gamma\, \exp[-\beta{\cal H}(\Gamma;A)]}
\approx
\frac{\int dx\, \exp[-\beta H(x;B)]}
     {\int dx\, \exp[-\beta H(x;A)]}
=
\frac{Z_B}{Z_A} = e^{-\beta\Delta F},
\end{equation}
where $Z_\lambda = \int dx\,\exp[-\beta H(x;\lambda)]$ is 
the appropriate expression for the partition function of the system of interest
when its coupling to the environment is negligibly weak.
\end{thebibliography}
\end{document}